\documentclass[12pt,a4paper]{article}
  \usepackage{a4wide}
  \usepackage{latexsym}
  \usepackage{epsf}
  \usepackage{amssymb}
  \usepackage{graphicx}
  \usepackage{amsmath, cite}
  \usepackage{amsmath,amssymb,amsthm}
  \usepackage{verbatim}
  \usepackage{hyperref}
  \usepackage{color}
  \usepackage{mathtools}
  \usepackage{soul,xcolor}

\renewcommand{\d}{\textrm{d}}

\newcommand{\w}{\wedge}

\newcommand{\rmi}{{\rm i}}

\DeclarePairedDelimiter\abs{\lvert}{\rvert}%
\usepackage{makecell}

\newcommand{\be}{\begin{equation}}
\newcommand{\ee}{\end{equation}}

\begin{document}
\numberwithin{equation}{section}

\vspace{0.4cm}
\begin{center}

{\LARGE \bf{Comments on classical AdS flux vacua\\ \vspace{0.4cm} with scale separation}}

\vspace{2 cm} {\large Fien Apers$^a$, Miguel Montero$^b$, Thomas Van Riet$^{c}$, Timm Wrase$^{d,e}$}\\
\vspace{1 cm} {\small\slshape $^a$Rudolf Peierls Centre for Theoretical Physics
Beecroft Building, \\Clarendon Laboratory, Parks Road, University of Oxford, OX1 3PU, UK}\\
\vspace{0.2 cm} {\small\slshape $^b$ Jefferson Physical Laboratory, Harvard University, \\ 
17 Oxford st Cambridge, MA 02138, USA}\\
\vspace{0.2 cm} {\small\slshape $^c$Instituut voor Theoretische Fysica, K.U.Leuven, \\Celestijnenlaan 200D, B-3001 Leuven, Belgium}\\
\vspace{0.2 cm} {\small\slshape $^d$Institute for Theoretical Physics, TU Wien, \\ Wiedner Hauptstrasse 8-10/136, A-1040 Vienna, Austria}\\
\vspace{0.2 cm} {\small\slshape $^e$Department of Physics, Lehigh University,\\ 16 Memorial Drive East, Bethlehem, PA 18018, USA}
\footnote{ fien.apers \emph{at} physics.ox.ac.uk, mmontero \emph{at} g.harvard.edu , Thomas.VanRiet \emph{at} kuleuven.be, Timm.Wrase \emph{at} lehigh.edu }


\vspace{1cm}

{\bf Abstract} \end{center} {\small AdS flux vacua with a parametric separation between the AdS and KK scales have been conjectured to be in the Swampland. We study flux compactifications of massive IIA supergravity with O6 planes which are claimed to allow moduli-stabilised and scale separated AdS$_3$ and AdS$_4$ vacua at arbitrary weak coupling and large volume. A recent refinement of the AdS Distance Conjecture is shown to be inconsistent with this class of AdS$_3$ vacua because the requisite discrete higher form symmetries are absent. We further perform a tree-level study of non-perturbative decays for the nonsupersymmetric versions of the AdS$_3$ solutions, and find that the vacua are stable within this approximation. Finally, we provide an initial investigation of the would-be dual CFT$_2$s and CFT$_3$s. We study roughly a dozen different models and find for all AdS$_4$ DGKT-type vacua that the dual operators to the lightest scalars have integer dimensions. For the putative CFT$_2$ dual theories of the AdS$_3$ vacua we find no integer dimensions for the operators.}
\newpage
\section{Introduction}
Perhaps the most elementary property of a vacuum solution of string theory, required for phenomenology, is that the extra dimensions are small enough to hide them from low-energy observers. In the case of AdS or dS vacua we can define ``small enough'' with respect to the Hubble scale. In this paper we focus on AdS vacua, and this condition reads
\be\label{scalesep}
\frac{L_{\text{KK}}}{L_{AdS}}\ll 1\,,
\ee
where $L_{\text{KK}}$ is the Kaluza-Klein lenghth scale defined through the overall volume $\mathcal{V}$ of the $d$-dimensional compact internal manifold as $\mathcal{V}=L_{\text{KK}}^d$ and $L_{AdS}$ is the inverse of the Hubble scale. We use the volume as a proxy for the masses of the tower of Kaluza-Klein modes of the internal space, which is a good idea if the internal  manifold is approximately isotropic. In general we want the masses of Kaluza-Klein excitations to be high with respect to the inverse AdS length. 

Another crucial property for phenomenology is moduli stabilisation. In case the AdS vacuum should be suitable for further uplift to a dS vacuum, we also want all squared masses to be positive, and not just above the BF bound. Such AdS vacua are really hard to come by, especially if we want parametric control and parametric scale separation\footnote{It has been claimed that orientifolds are necessary for achieving this \cite{Gautason:2015tig}, but recent investigations \cite{DeLuca:2021mcj, Cribiori:2021djm} might imply this requirement can be dropped.}. A holographic perspective seems to point to the same difficulty. Scale separation with positive masses implies dead-end CFTs (that is, CFTs without any relevant or marginal deformations) with parametric gaps in the operator spectrum. Such CFTs have never been constructed before, but neither is there a proof against their existence. A recent analysis of \cite{Collins:2022nux} however supports the intuitive picture that such CFTs potentially cannot exist and the authors suggest that the first non-trivial spin two operator of a CFT (dual to a KK mode) cannot have a parametric large dimension, similar in spirit to the conjecture made earlier for spin zero operators \cite{Gautason:2018gln}.\footnote{The AdS moduli conjecture of \cite{Gautason:2018gln} is still consistent with scale separated vacua of the DGKT kind.}

We are aware of two classes of flux compactifications which are claimed to nevertheless achieve the above mentioned properties.  Most well-known are the so called DGKT AdS$_4$ vacua \cite{DeWolfe:2005uu} (see also \cite{Derendinger:2004jn,Behrndt:2004mj,Camara:2005dc}) from reducing massive IIA supergravity on a CY 3-fold with intersecting O6 planes down to four dimensions using RR and NSNS fluxes.  Preliminary investigations on finding solutions in IIB, inspired from T-duality, exist \cite{Caviezel:2009tu, Petrini:2013ika} but they are outside of the controlled regime since some cycles become small \cite{Cribiori:2021djm}. However, double-T-duality brings one back to IIA without Romans mass and then controlled solutions seem again possible \cite{Caviezel:2008ik, Cribiori:2021djm}. A second class consists of AdS$_3$ vacua obtained by reducing massive IIA on a manifold of G2 holonomy \cite{Farakos:2020phe, Emelin:2021gzx}. For both classes of examples a certain set of RR fluxes are unconstrained by tadpole conditions and taking a limit of large flux guarantees parametric weak coupling, large volumes, and parametric scale separation. 

One of our results is a basic investigation of the putative CFT$_2$ dual to the AdS$_3$ vacua, akin to the one carried out for the DGKT AdS$_4$ vacua in \cite{Aharony:2008wz, Conlon:2021cjk}. We will also extend the analysis of AdS$_4$ vacua in \cite{Conlon:2021cjk} by looking at many different toroidal orbifolds, more general examples with metric fluxes and a dual type IIB compactification. An important focus in this paper is whether the AdS$_3$ vacua satisfy some conjectured Swampland criteria. We focus on three conjectures, relevant to our discussion:
\begin{itemize}
    \item The Strong AdS distance conjecture \cite{Lust:2019zwm} ruling out all scale separated SUSY AdS vacua and its refined version \cite{Buratti:2020kda} that rules in the DGKT vacua on the account of specific properties related to the presence of a discrete higher form symmetry, which we review.
    \item The non-SUSY AdS conjecture \cite{Ooguri:2016pdq} stating that all non-SUSY AdS spaces can at best be meta-stable. 
    \item The AdS moduli conjecture \cite{Gautason:2018gln} stating that the lightest scalar mass $m$ should not be parametrically large in AdS units, i.e., $m^2L_{AdS}^2$ is not parametrically large.
\end{itemize}
The last conjecture is automatically satisfied for the AdS$_3$ vacua of \cite{Farakos:2020phe} and the AdS$_4$ vacua of \cite{DeWolfe:2005uu}. Therefore we focus on the first two conjectures. We will find the first one is violated in the AdS$_3$ compactifications of \cite{Farakos:2020phe},  and that the decay channel required by the second one cannot be found, within the approximations made. 

The reason we check the consistency of the AdS flux vacua with Swampland conjectures instead of using the vacua to give circumstantial evidence or counterexamples to said conjectures is that the vacua have not been established at the full string theory level. Their existence is still shrouded in some mystery due to approximations used in deriving the vacua.  For instance, for the AdS$_3$ vacua the O6 planes are distributed as follows on the toroidal covering space:
\begin{align}
\begin{pmatrix} 
&{\rm O}6_{\alpha}:\quad & \times & \times & \times & \times & - & - & -  \\
&{\rm O}6_{\beta} :\quad &\times & \times & - & - & \times & \times & -  \\
&{\rm O}6_{\gamma}:\quad &\times & - & \times & - & \times & -& \times  \\
&{\rm O}6_{\alpha\beta}:\quad & - & - & \times & \times & \times & \times & -  \\ 
&{\rm O}6_{\beta\gamma} :\quad & - & \times & \times & - & - & \times & \times  \\ 
&{\rm O}6_{\gamma\alpha} :\quad &- & \times & - & \times & \times & - & \times  \\ 
&{\rm O}6_{\alpha\beta\gamma}:\quad & \times & - & - & \times & - & \times & \times 
\end{pmatrix} \, . \label{intersection}
\end{align}
Such a complicated intersection of O6 planes hinders a clear 10-dimensional picture in which the O6 planes backreact on the G2 holonomy \cite{Banks:2006hg, McOrist:2012yc}. The only available 10D picture is one in which the O6 planes are smeared over the G2 space \cite{Farakos:2020phe}, just as for the DGKT vacua \cite{Acharya:2006ne, Grana:2006kf, Caviezel:2008ik}. In the recent years this has been understood better, and for non-intersecting O6 planes in massive IIA the explicit backreaction was shown to be consistent with the smeared approximation \cite{Baines:2020dmu} whereas for the intersecting case reassuring results were obtained at first-order in a backreaction series \cite{Junghans:2020acz, Marchesano:2020qvg}. A generalisation of the DGKT solutions without Romans mass exists, and a preliminary lift to 11d has not revealed any signs of a troublesome backreaction either \cite{Cribiori:2021djm}, despite the claims in \cite{Banks:2006hg}. For the AdS$_3$ vacua of \cite{Farakos:2020phe} this has not yet been achieved but some first step in generalising them was taken in \cite{Emelin:2021gzx}.

In  the next section we review the flux vacua of \cite{Farakos:2020phe} and elucidate the vector spectrum which was not known. In section \ref{sec:dis} we verify that the refined Strong AdS distance conjecture of \cite{Buratti:2020kda} is violated despite the close analogy with the DGKT vacua. We also tried to construct explicit non-perturbative decay channels for the non-SUSY AdS$_3$ solutions of \cite{Farakos:2020phe}, and found none,  paralleling the analysis carried out for the the DGKT vacua in \cite{Narayan:2010em} (see however \cite{Marchesano:2021ycx}, which achieved recent progress in DGKT). Finally, in section \ref{sec:conlon} we compute some basic properties of the dual CFTs for AdS$_3$ and AdS$_4$ vacua and find interesting results for the dimensions of dual conformal operators. Much of the work in this paper that concerns AdS$_3$ vacua is based on the master's thesis \cite{Fien}.

\section{Review of scale separated AdS$_3$ vacua}\label{sec:rev}
We now recall the construction of scale separated AdS$_3$ vacua of \cite{Farakos:2020phe} and extend the discussion of the spectrum of light fields to the case of vectors. The procedure of \cite{Farakos:2020phe} mimics essentially the construction of the DGKT AdS$_4$ vacua \cite{DeWolfe:2005uu} (see also \cite{Derendinger:2004jn,Behrndt:2004mj,Camara:2005dc}) which were obtained from reducing massive IIA supergravity on a CY 3-fold with intersecting\footnote{Whether the O6 planes intersect depends on the details of the orbifold. In the original example of \cite{DeWolfe:2005uu} they do not intersect.} O6 planes down to four dimensions. To obtain AdS$_3$ vacua we instead reduce on a manifold of G2 holonomy. A certain set of 4-form RR fluxes are unconstrained by tadpole conditions and taking a limit of large flux guarantees parametric weak coupling, large volumes, and parametric scale separation.

To make the construction explicit we will orientifold an orbifolded 7-torus such that we obtain a singular G2 space. The seven internal coordinates on the torus covering space are labeled as $y^m$ 
\be
y^m \simeq y^m+1 \, ,
\ee
and we take the metric in Einstein frame to be
\begin{equation}ds^2_{\text{Covering $T^7$}}= \sum_m (r^m)^2 (dy^m)^2.\end{equation}
The orbifold group $\Gamma$ is generated by the following $\mathbb{Z}_2$ involutions 
\be
\begin{aligned}
	\Theta_\alpha : (y^1, \dots, y^7 ) & \to (-y^1, -y^2, -y^3, -y^4, y^5, y^6, y^7) \, , 
	\\
	\Theta_\beta : (y^1, \dots, y^7 ) & \to (-y^1, -y^2, y^3, y^4, -y^5, -y^6, y^7) \, ,
	\\
	\Theta_\gamma : (y^1, \dots, y^7 ) & \to (-y^1, y^2, -y^3, y^4, -y^5, y^6, -y^7) \, , 
\end{aligned}
\ee
and then $\Gamma=\{\Theta_\alpha,\Theta_\beta,\Theta_\gamma, \Theta_\alpha\Theta_\beta, \Theta_\beta \Theta_\gamma, \Theta_\gamma \Theta_\alpha, \Theta_\alpha \Theta_\beta \Theta_\gamma\}$. 
The $\Theta$'s commute and preserve the calibration three-form $\Phi$: \be
\Phi = e^{127} - e^{347} - e^{567} + e^{136} - e^{235} + e^{145} + e^{246} \, , 
\ee 
where $e^{ijk} = e^i \wedge e^j \wedge e^k$, and we have introduced the seven vielbeins of the torus
\be
e^m = r^m dy^m \, . 
\ee
The untwisted sector is described by a compactification over a singular G2 space without 1- or 6-cycles, and with seven 3-cycles (and seven dual 4-cycles). In general, we would expect additional fields in the twisted sector; some of these might be interpreted as collapsed 2- or 5-cycles. In any case, the untwisted sector  gives eight scalars (the 7 radii $r^m$ and the dilaton). Reducing $C_3$ over the 2-cycles leads to 3D vectors which we will show below are all massive, if present. The $C_1$ vector is projected out, so the whole untwisted bosonic content comprises of eight real scalars. The fluxes stabilise these 8 scalars.  

\subsection{Fluxes and orientifolds}
We add O2 planes through the following involution (together with the usual worldsheet actions): \be
\sigma : (y^1, \dots, y^7 )  \to (-y^1, -y^2, -y^3, -y^4, -y^5, -y^6, -y^7) \, . 
\ee  
This involution $\sigma$ has $2^7$ fixed points, and thus $2^7$ different O2 sources in the torus covering space located at the points $y^i=0, 1/2$.  Notice that the calibration is odd under the O2 involution 
and that $\Gamma$ commutes with $\sigma$.  The orbifold images of these O2 planes are O6 planes specified by the involutions:
\begin{align}
\Theta_{\alpha}\sigma : y^i & \to (y^1, y^2, y^3, y^4, -y^5, -y^6, -y^7) \,, \nonumber\\
\Theta_{\beta} \sigma : y^i & \to (y^1, y^2, -y^3, -y^4, y^5, y^6, -y^7)\,,\nonumber\\
\Theta_{\gamma}\sigma : y^i & \to  (y^1, -y^2, y^3, -y^4, y^5, -y^6, y^7)\,,
\end{align}
and products thereof. This leads to 7 different directions for O6-planes already depicted in equation \eqref{intersection}. These intersections are calibrated supersymmetrically and their transversal spaces are defined by the volume forms
\be
\label{3basis}
\Phi_i = \left( dy^{127}, - dy^{347}, - dy^{567}, dy^{136}, - dy^{235}, dy^{145}, dy^{246} \right) \ , \quad i = 1 , \dots, 7 \, , 
\ee
which form a useful basis of 3-forms. The G2 calibration is then $\Phi = s^i \Phi_i$,  where the $s^i$ are the metric moduli, related to the radii $r^m$ as follows 
\be
s^1 \Phi_1 = e^{127} \ \to \ s^1 = r^1 r^2 r^7 \ , \quad  s^2 \Phi_2 = -e^{347} \ \to \ s^2 = r^3 r^4 r^7 \ , \quad  \text{etc.} 
\ee 
A basis of (co-)closed 4-forms, invariant under $\Gamma$ is
\be
\label{4basis}
\Psi_i = \left( dy^{3456}, -dy^{1256}, - dy^{1234}, dy^{2457}, - dy^{1467}, dy^{2367}, dy^{1357} \right) \ , \quad i = 1 , \dots, 7 \, . 
\ee
The co-associative calibration takes the form 
\be
\label{starex}
\star \Phi = \sum_{i=1}^{7} \frac{V_E}{s^i} \Psi_i \, ,
\ee
where the volume in Einstein frame is given by $V_E=r^1 r^2 \ldots r^7$.

We now specify the fluxes; we will have NSNS $H_3$ flux and the RR fluxes $F_4$ and $F_0$. The $F_4$ flux splits into two parts:
\be
F_4 = F_{4A} + F_{4B}\,,
\ee
with the split determined by its wedge product with $H_3$:
\be
[F_{4A}\wedge H_3] = 0 \,,\qquad [F_{4B}\wedge H_3] = -[\delta_{O2}],
\ee
where $[X]$ denotes the de Rham cohomology class of the differential form $X$. The actual 10-dimensional equations require equality of the flux at the level of differential forms; restricting ourselves to solving them at the level of cohomology classes only is equivalent to smearing the $O$-planes.  We see that the $B$-part is designed to cancel the O2 RR tadpole, while the A-part is unconstrained by any tadpole. This will turn out crucial for achieving weak coupling, large volume, and scale separation. The O6 tadpole is cancelled by $F_0$ and $H_3$ flux. The 10D solution (and hence with smeared orientifolds) is given by
\be
H_3 = h\sum_i \Phi_i \,,\qquad F_{4A} =\sum_i f^i\Psi_i \,,\qquad  F_{4B} =\sum_i \hat{f}^i\Psi_i\,.
\ee
Tadpole constraints enforce:
\be
hm=\mu\,, \qquad \sum f_i=0\,,\qquad \sum \hat{f}_i=\frac{\mu'}{h}\,.
\ee
where $F_0=m$, $\mu$ is the O6-plane charge and $\mu'$ the O2-plane charge.

\subsection{Stabilisation of scalars}
The 3D supergravity theory has minimal supersymmetry in the form of two real supercharges. The 3D scalar potential is determined by a real superpotential $P$ via
\be\label{Pfunc}
V(\phi) = G^{IJ}  P_I P_J  - 4 P^2 \, . 
\ee 
$P_I$ is shorthand for $\partial_I P$ and $G_{IJ}$ is the metric on the scalar manifold. In \cite{Farakos:2020phe} the general expression for $P$ for any G2 compactification with fluxes and O2/O6 planes was found. We do not need it here, and just present the real superpotential $P$ for the simple fluxes given earlier: 
\be
\label{SPSP-TOT}
P = \frac{m}{8} e^{\frac{y}{2}  - \frac{\sqrt 7 x}{2}} 
+ \frac{ h}{8} e^{y + \frac{x}{\sqrt{7}}} \sum_{i=1}^7 \frac{1}{\tilde s^i} 
+  \frac{1}{8} e^{y - \frac{x}{\sqrt{7}}} \sum_{i=1}^7  ( f^i + \hat f^i ) \tilde s^i 
\ , \quad 
\tilde s^7 = \prod_{a=1}^{6} \frac{1}{\tilde s^a} \, . 
\ee
We defined orthogonal combinations $x,y$ of the dilaton $\phi$ and volume modulus $v$
\begin{equation}
    \dfrac{x}{\sqrt{7}} = -\dfrac{3\phi}{8} + \dfrac{\beta v}{2}, \quad 2y = -21 \beta v -\dfrac{1}{4} \phi,
\end{equation}
where the Einstein frame volume is defined by $V_E=\exp(7\beta v)$ in string units and 
\begin{equation}
 \tilde{s}^a = V_E^{3/7} s^a\,. 
\end{equation}
Canonical normalisation of $x, y$ (or $v,\phi$) requires $\beta=\frac{1}{4\sqrt{7}}$.

A particularly simple solution has the following choices where $F_{4A} \ne 0$ and $F_{4B}=0$: 
\be
\label{fluxesANSA}
\hat f^i = 0 \ , \quad f^i = (-f,-f,-f,-f,-f,-f,+6f)   \, , \quad f\,,h\,,m>0 \,. 
\ee
Since there is no $F_{4B}$ we require space-filling D2-branes to solve the RR tadpole, and these introduce many additional compact moduli. However, as shown in \cite{Farakos:2020phe} introducing the right amount of $F_{4B}$-flux to cancel the O2 tapdole without D2 branes still leads, parametrically, to the vacuum described with the above fluxes in the large $f$ limit. This is because the number of $D2$ branes one needs to introduce to cancel the tadpole does not scale with $f$. Hence the above vacuum in the large $f$ limit serves as some universal solution, good for understanding the closed string untwisted sector, whose properties we now outline. Later we use this vacuum as a starting point for studying its putative holographic dual.

In total we need to stabilise 8 scalars; the dilaton $\phi$, the internal volume in 10D Einstein frame $V_E$, and six fluctuations of radii at fixed volume $\tilde{s}^a$ with $a=1, \ldots, 6$. One should worry about the twisted sector which we ignore. However, one expects that this is not a genuine problem since it was explicitly verified that the analogous issue does not arise in the DGKT vacuum \cite{DeWolfe:2005uu}, although this issue is currently under investigation \cite{progress}.

The simplest supersymmetric AdS vacua considered in \cite{Farakos:2020phe} assumed all $\tilde{s}^a$ ($a=1\ldots 6$) have the same value $\sigma$: 
\be
\langle \tilde s^a \rangle = \sigma\,. 
\ee
The resulting SUSY vacuum has:
\begin{align}
\label{aANDb}
& \frac{h}{f} V_E^{1/7}e^{-3\phi/4} =a \ , \quad  \frac{m}{f} e^{\phi}V_E^{4/7} = b \,,\\
& \langle V \rangle = - \frac{1}{64 a^6 b^4} \left( 6 \sigma^2 + \frac{36}{\sigma^{12}} \right)  \frac{m^4 h^6}{f^8}   \, ,  
\end{align}
with 
\be
\label{numera}
a = 0.515696\dots \ , \quad b =  3.43111\dots \ , \quad \sigma = 1.32691\dots \, . 
\ee

\subsection{Scale separation}
The flux $f$ is unconstrained by tadpoles and can be taken arbitrary large. In the large $f$ limit we find:
\be
g_s = e^{\phi} \sim f^{-3/4} 
\ , \quad 
V_E \sim f^{49/16} \, , 
\ee
and we can thus verify that our vacuum corresponds to weak string coupling and to large volume for large $f$. The volume in 10D string frame $V_S$ scales as $V_S \sim f^{7/4}$ which also grows large. The AdS radius scales as $L^{-1}_{AdS}\sim M_p \sqrt{\langle V\rangle }$ such that we find arbitrary large scale separation\footnote{$M_p$ in 3D is given by $g_s^{-2}V_S$.}
\be
\frac{(V_S)^{1/7}}{L_{AdS}}  \sim f^{-1/2} \, , 
\ee
at arbitrary \emph{small coupling and large volume}. In this solution, all of the six $\tilde s^a$ take the same numerical value, $\sigma$, by construction, and the seventh is slightly different. But the torus remains, as a whole, at large values and no individual directions get small.

\subsection{Axion content}
Reference \cite{Farakos:2020phe} did not investigate the axion content of the vacua and so we do this here and demonstrate that the fluxes also remove all axions. Axions can either come from D-brane positions, from dualising vectors on the D-brane worldvolume, from dualising vectors from reducing $C_3$ over even two cycles or from reducing $B_2$ over odd 2-cycles. Since we can cancel RR tadpoles with fluxes, no D-branes are needed. 

The only two cycles in our model are potentially in the \emph{twisted sector}, if any. We will now show we expect the vectors generated from $C_3$ to be massive and so no axions are being generated by $C_3$. Whether axions from $B_2$ can generate discrete symmetries is discussed below. 

The Chern-Simons term in 10D that can give these vectors a  mass upon reduction, is 
\be
S_{10} = - \frac12 \int_{10} C_3 \wedge H_3^{\text{BG}} \wedge dC_3 \, .  
\ee
If we insert 
\be
H_3^{\text{BG}} = \sum_{i=1}^{b_3} h^i \Phi_i \ , \quad C_3 = \sum_{a=1}^{b_2} A^a_\mu \, dx^\mu \wedge \omega_a \, .  
\ee
We then find 
\be
{\cal L}_3 = - \frac12 m_{ab}  \, A^a \wedge d A^b \ , \quad m_{ab}=  \lambda_{abj} \, h^j \, , 
\ee
where $a,b=1, \dots, b_2$ (betti-number for even $2$-cyles) and $j=1,\dots,b_3$ and  
\be
\int_7 \omega_a \wedge \omega_b \wedge \Phi_j = \lambda_{abj} \, . 
\ee
If we use that the flux vacuum has all $h_i$ equal; $h_i=h$, and that we can choose calibrated representatives for the basis of 2-cycles\footnote{Such that $\star \omega = \omega \wedge \Phi$.} we find
\be
m_{ab} \sim h\int \star \omega_a \wedge \omega_b\,,
\ee
which is clearly full rank.

\section{Conjectured Swampland criteria}\label{sec:dis}

The refined strong AdS distance conjecture \cite{Buratti:2020kda} allows scale separated AdS vacua on the condition that there is a large discrete higher form symmetry with some specific properties and it was shown that the DGKT vacua \cite{DeWolfe:2005uu} obey those properties. In here we investigate whether the same is true for the AdS$_3$ vacua of \cite{Farakos:2020phe}. We first briefly summarize the findings of \cite{Buratti:2020kda} concerning the scale separated AdS$_4$ DGKT vacua.

\subsection{$\mathbb{Z}_k$ weak coupling and refined AdS distance conjecture}
Reference \cite{Buratti:2020kda} considers a Gedanken experiment involving a charged Reissner-Nordstr\"om black hole that is also charged under a discrete $\mathbb{Z}_k$ symmetry, when $k$ becomes large. The $\mathbb{Z}_k$ gauge symmetry does not involve a local field, and therefore the semiclassical space-time solution for the black hole is blind to the $\mathbb{Z}_k$-charge.  Following a logic similar to that used to derive the species bound \cite{Dvali:2007hz}, one arrives at a condition for the masses of the lightest $\mathbb{Z}_k$-charged particles:
\be\label{species}
m\sim k^{-\alpha}M_p\,,
\ee
where $\alpha$ is positive and of order 1.  In case of a supersymmetric theory with extremal non-SUSY black holes, the most natural thing the black hole can do in order to decay consistently is making sure the WGC-satured particles can carry away the $\mathbb{Z}_k$-charge as well. Since those particles satisfy
\be
m\sim gqM_p\,,
\ee
the most natural way this is consistent is if $g\sim k^{-\alpha}$.  This is called the $\mathbb{Z}_k$ Weak Coupling Conjecture (WCC).

It is important to realize that, in the setup considered in \cite{Buratti:2020kda},  the $\mathbb{Z}_k$ charge is correlated with the U(1) charge. The $\mathbb{Z}_k$ charged particles that are emitted in the evaporation process are neutral under an additional $\mathbb{Z}_m$ gauge symmetry, where $k \gg m$. The $U(1)$ generator is in fact a linear combination of the $\mathbb{Z}_k$ and the $\mathbb{Z}_m$ generators. The $\mathbb{Z}_m$ charged particles satisfy the $\mathbb{Z}_m$ WGC and are thus heavier, in a natural realization of a sublattice WGC \cite{Harlow:2022gzl}.
Reference \cite{Buratti:2020kda} then took the natural next step and contemplated the same conditions for higher-form gauge symmetries under which extended objects can have both $\mathbb{Z}_k$ and $U(1)$-charges. These ideas were then tested in various circumstances and our interest is particularly in the situation of the DGKT vacua. Such vacua were shown to have a $\mathbb{Z}_k$ 3-form symmetry under which domain walls are charged and at the same time they have the usual charge under a massless 3-form. 

A very intriguing observation was then made concerning scale separation. Recall that the Strong AdS Distance Conjecture (SADC) \cite{Lust:2019zwm} states that for supersymmetric AdS vacua 
\be
\Lambda \sim M^2_{\text{cutoff}} \,.
\ee
Often $M_{\text{cutoff}}$ means $M_{\text{KK}}$, and then the SADC clearly forbids scale separation. If this is correct, the DGKT vacua must either be inconsistent, or the cutoff scale is unexpectedly low.

The relation between scale separation and the $\mathbb{Z}_k$ WGC comes about as follows; the fluxes taken large in DGKT with quantum $k$ indeed lead to a discrete $\mathbb{Z}_k$ symmetry on top of a continuous one. The WGC domain walls are charged both under this discrete  and continuous symmetry, and are unstable against nucleation of strings. Using the interplay of these $U(1)$ and discrete symmetries, the authors of \cite{Buratti:2020kda} could derive how the internal volume and the dilaton, and hence also $M_{KK}$, depend on $k$. Furthermore, there are stable domain walls, not charged under any discrete symmetry, that separate vacua with a different flux number $k$. Their tension directly relates to the vacuum energy through a relation of the kind $\d \sqrt{|\Lambda|}/\d k \sim T_{DW}$, in Planck units, with the tension scaling in a certain way with $k$. This equation essentially comes from the junction condition for domain walls that make $k$ jump one unit.   As a consequence one can argue for the relation, similar to the black hole case \eqref{species} with $\alpha=1$:
\be\label{rSADC}
\Lambda \sim \frac{M^2_{\text{KK}}}{k} \,.
\ee

For large $k$ there is indeed scale separation. This is how \cite{Buratti:2020kda} came to the refined SADC, which essentially states that in a SUSY  AdS$_d\times X$ vacuum, with domain walls charged under a continuous and a discrete $\mathbb{Z}_k$ $(d-1)$-form symmetry, \eqref{rSADC} has to hold, implying scale separation is possible at large $k$.  Note that the heuristic derivation is rather loose and one should regard \eqref{rSADC} mainly as an observation for the IIA scale separated AdS$_4$ vacua that looks very much like the domain wall extension of \eqref{species}.

We now review how the $\mathbb{Z}_k$  $3$-form symmetry arises in the DGKT vacua and then show that the analogous $\mathbb{Z}_k$  $2$-form symmetry is unfortunately absent for the AdS$_3$ vacua of \cite{Farakos:2020phe}. This shows that the refined SADC, at least in the form proposed in \cite{Buratti:2020kda}, is not applicable in 3 dimensions. 

\subsection{Discrete symmetries and AdS$_4$ vacua}

The discrete 3-form symmetries of DGKT vacua follow from reduction of the $F_4F_4B_2$ CS-term in 10D. When reduced to 4D, the internal $F_4$ integral provides the factor of $k$ and the $B_2$ integral over internal dual 2-cycles gives a linear combination of axions $\chi_i$. One then arrives at the following term in the 4D action:
\be\label{Kaloper}
k (\sum \chi_i) F_4\,.
\ee
Through the Kaloper-Sorbo-Dvali mechanism \cite{Kaloper:2008fb, Dvali:2005an} this gives the 3-form a mass by eating the two-form dual to $\sum \chi_i$. This mass means there is effectively a discrete 3-form $\mathbb{Z}_k$ symmetry. Reducing also the $F_0B_2F_8$ makes the story somewhat more complicated since the $F_8$ reduced over 4-cycles gives extra 4-form field strengths coupling to the axions. But the overall factor goes like $F_0=m$ and does not scale up with the $F_4$ flux. This mixing implies there is a massless and a massive combination of 3-forms and accordingly a continuous and a discrete 3-form symmetry. 
Relatedly, we have stable and unstable domain walls; a certain  combination of space-filling D2 branes and D6 branes wrapping 4-cycles are unstable, whereas D4 branes wrapping holomorphic 2-cycles are SUSY and stable. Reference \cite{Buratti:2020kda} verified that their couplings and tensions obey the discrete WGC and WCC. The D4 brane tension relates to the vacuum energy and reproduces the rSADC \eqref{rSADC}.  

\begin{figure}[!h]
	\centering
	\includegraphics[width=0.45\textwidth]{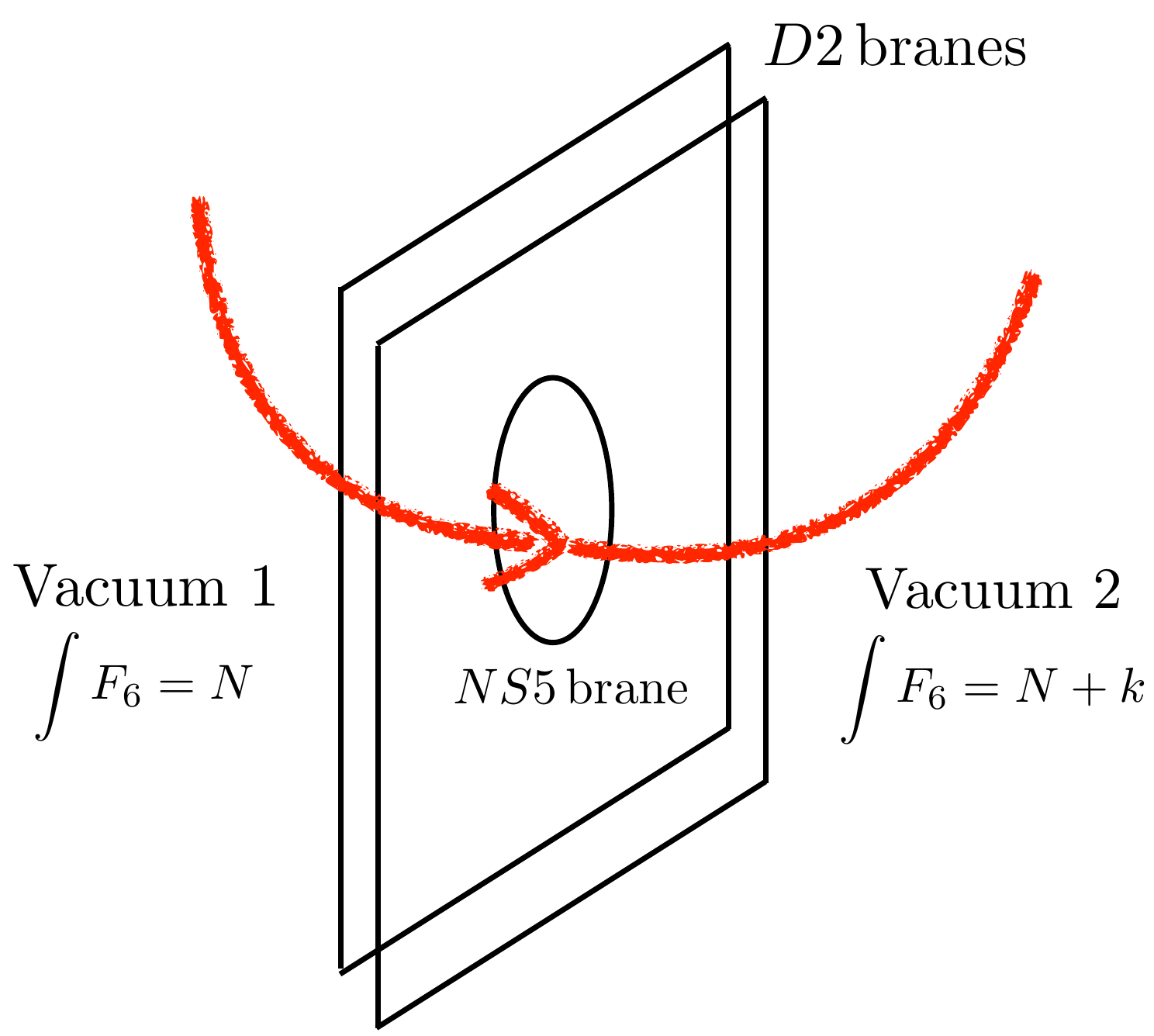}
	\caption{\emph{The unstable D2 membrane nucleates holes in it whose boundaries are wrapped NS5 branes looking like closed strings in 4D.}} \label{Figure}
\end{figure}

The instability of the D2 and D6 branes occurs through the nucleation of holes inside of the walls, as depicted in Figure \ref{Figure}.
The boundaries of these holes are formed by closed strings in 4D, which descend from NS5 branes wrapped over 4-cycles. These are exactly the branes that couple to the axions obtained from reducing $B_2$ over the dual 2-cycles. One way to think of this process is through Freed-Witten anomalies \cite{Freed:1999vc}: The RR flux through these 4-cycles causes the anomalies implying the NS5 branes ``emit" the D2 and D6 branes. From a 4D viewpoint this process is really the appearance of holes in the non-BPS combination of these domain walls, whose instability follows from the mass-term induced by the above coupling \eqref{Kaloper}.

\subsection{Discrete symmetries and AdS$_3$ vacua}
Here we search for discrete 2-form symmetries and stable and unstable domain walls in 3D (which are 1+1 dimensional objects in this case). The flux-parameter taken large in the AdS$_3$ solutions is $F_4$ flux, which we denote as $f$ before. It is integer quantized, and by analogy with the case above, we will search for a discrete $\mathbb{Z}_f$ symmetry, possibly correlated with a continuous 2-form symmetry.  To do that, we first go through the 3-form field strengths systematically, taking into account orientifold projections \footnote{Under the O2/O6 projection we have that $H_3, F_2, F_6$ are odd while $F_0, F_4, F_8$ are even.}. The fields they yield in three dimensions after dimensional reduction are summarized in the following table:
\begin{center}
\begin{tabular}{c|c|c|c|c|c|c}
10d field&point&2-cycle&3-cycle&4-cycle&5-cycle&Total space\\\hline
$H_3$&--&Axion&Flux&--&--&--\\
$H_7$&--&--&--&2-form&Vector&--\\
$F_0$&Flux&--&--&--&--&--\\
$F_{10}$&--&--&--&--&--&2-form\\
$F_{2}$&--&Flux&--&--&--&--\\
$F_{8}$&--&--&--&--&2-form&--\\
$F_{4}$&--&Vector&Axion&Flux&--&--\\
$F_{6}$&--&--&2-form&Vector&Axion&--
\end{tabular}\end{center}
Entries corresponding to 2-cycles and 5-cycles can only arise from twisted sector fields. There are no (non-torsion) 1-cycles or 6-cycles in a $G_2$ manifold, and for the two-, three- and four-cycle entries one must choose even or odd representatives under the orientifold actions as dictated by the parity of the 10d fields. In 3d, vectors may be dualized to scalars if they are not appearing in Chern-Simons terms. Note that the table overcounts the field content because of Hodge duality in 10D, which translates to Hodge duality in 3D\footnote{Since the Hodge duality in 3D can be involved because of St\"uckelberg couplings.}; for instance a 2-form in 3D is flux, etc. Hence, it suffices to analyse the fields generated by $H_7, F_6, F_8$ and $F_{10}$. We first investigate the possible 2-forms needed for Davli-Kaloper-Sorbo (DKS) couplings:  
 \begin{itemize}
 	\item The space-filling $H_3$ fieldstrength is projected out by the orientifold planes. $H_7$ reduced on even 4 cycles gives 3-form fieldstrengths. There is no $H_7C_3$ CS term (or $B_6 F_4$ term), hence the associated 2-forms are massless. 
 	\item  We can reduce $F_6$ over odd 3-cycles. The CS terms that could give a mass are $F_6F_2B_2$. But there is no $F_2$ flux, nor axions from $C_1$. So the corresponding 2-forms are massless.
 	\item We can reduce $F_8$ over even 5 cycles (if any) to obtain a 3-form fieldstrength. A relevant CS term could be $B_2F_8F_0$. With the axions from $B_2$ over odd 2-cycles this gives a mass of order $m$, and so a $\mathbb{Z}_m$ discrete symmetry; but there is no discrete symmetry whose order grows with $f$. 
 	\item The RR potential $C_9$ for Roman's mass can be integrated over the whole internal space space. This is clearly a massless 2-form. The charged objects are D8 branes wrapped on the internal manifold, and they automatically satisfy the WGC. 
 \end{itemize}
 So, unlike in the previous case, we do not see any 2-form field (with 3-form strength) that could get a mass of order $f$. 
 
 Alternatively we could look for the axions needed in the DKS couplings. Vectors coming from $C_3$ receive a Chern-Simons mass, so we will remove their would-be dual axions from the count. Let us discuss for instance axions from $C_3$ reduced over even 3-cycles, or $B_2$ over odd 2-cycles. $C_3$ axions can only arise for even 3-cycles coming from twisted sectors, and those axions could only couple via the $C_3 F_4 H_3$ coupling to $H_3$ with all three legs in AdS$_3$, which is projected out. $B_2$ axions can only arise from 2-cycles in twisted sectors, and can couple to 3-form fieldstrengths arising from periods of $F_8$ on even 5-cycles, via the $F_0 B_2 F_8$ Chern-Simons term. We recover the $\mathbb{Z}_m$ discrete symmetry we found above, but none of order $f$. All in all, it seems we neither find massive 2-forms with masses of order $f$ nor the right kind of axions for the DKS couplings. 

For consistency we also verify whether we can see this as well from the brane structure and the possible Freed-Witten effects. These can be \cite{Herraez:2018vae}
\begin{enumerate}
\item A NS5 brane threaded by $F_p$ flux emits D$(6-p)$ branes.
\item A D$p$ brane threaded by $H_3$ flux emits D$(p-2)$'s.
\item A D$p$ brane threaded by $F_p$ flux emits fundamental strings. 
\end{enumerate}
A discrete $\mathbb{Z}_n$ symmetry is associated to a domain wall (a string in 2+1 dimensions) such that $n$ copies of it together can break. At the end-points we have particles. So the branes forming the particles at the end-points ``emit'' the branes corresponding to the unstable strings.  What can those particles be?
\begin{itemize}
    \item In analogy with \cite{Buratti:2020kda} the natural particle candidate here is an NS5 brane wrapping an even 5-cycle, which emits $D6$ branes wrapped on 5-cycles when threaded by $F_0=m$ units of flux. This is the string associated to the $\mathbb{Z}_m$ discrete symmetry uncovered above.
 \item D4 branes wrapping 4-cycles threaded by $F_4$ flux can emit fundamental strings, but these are projected out. Relatedly, one can verify that stable D4 cycles would be odd but then there is no Freed-Witten effect since $F_4$-flux is even.
 \item A D2 brane wrapping a 2-cycle threaded by $F_2$ flux; but there is no $F_2$ flux in the model (and no 2-cycles in the particular incarnation discussed in \cite{Farakos:2020phe}).
 \end{itemize}
 
 A similar conclusion is reached by a systematic analysis of the strings. We conclude that there is no $\mathbb{Z}_f$ 2-form symmetry, regardless of the content of the twisted sector. If the twisted sector contains even 5-cycles, there could be a discrete 2-form symmetry of order $m$.

\subsection{Non-SUSY AdS conjecture and fake supersymmetry} 
The SUSY AdS$_3$ vacua have almost identical non-SUSY sisters by flipping the sign of the four-form flux, which does not affect the mass spectrum and hence their perturbative stability \cite{Farakos:2020phe}.\footnote{Note that this is different from the DGKT model \cite{DeWolfe:2005uu} where non-SUSY solutions obtained by a sign flip of the $F_4$-flux quanta can have a different mass-spectrum, see subsection \ref{Timmsection}.} A natural question is then whether the non-SUSY AdS Swampland conjecture  \cite{Ooguri:2016pdq} holds. In other words, can we find a decay channel? Note that also the DGKT vacua have non-SUSY sisters, and the decay channels cannot be found in the thin-wall approximation at tree level, without including any quantum corrections for the tension, as shown in as shown in  \cite{Narayan:2010em} (see \cite{Marchesano:2021ycx} however for recent progress in instabilities of DGKT vacua). We now summarize those arguments when applied to the AdS$_3$ vacua.

In three dimensions the lower bound on domain wall tensions is given by
\begin{equation}\label{nablaP}
    T_L = 2 \abs{\nabla P},
\end{equation}
with $\Delta P$ the jump in the real 3D superpotential through the wall. The upper bound for decay to occur is given by
\begin{equation}
     T_U = \sqrt{\abs{V_{-}}}-\sqrt{\abs{V_{+}}}\,,
\end{equation}
with $V_{-}$ the potential of the true vacuum and $V_{+}$ the potential of the false vacuum. 

For the SUSY AdS$_3$ vacua we have $T_U = T_L$.  We can show this also happens for the  non-SUSY vacua using \emph{fake supersymmetry} \cite{Freedman:2003ax, DiazDorronsoro:2016rrz}: if we take our expression for $P$ and flip the sign of the last term (containing $F_4$ flux) then this `fake' $P$-function also solves the equation \eqref{Pfunc}, although it is not the true $P$-function seen in the SUSY variation of the gravitino. For this fake $P$ the non-SUSY AdS vacuum is in fact a critical point! And the domain wall solutions found from the gradient flow equations using the fake $P$ are therefore marginally stable \cite{Freedman:2003ax} and their tension is then given by \eqref{nablaP} for the fake $P$. However, on-shell the fake $P$ is identical to the on-shell real $P$ of the SUSY AdS vacuum since the sign flip in the P-function is compensated by them having opposite $F_4$ fluxes so both domain walls of the SUSY and the non-SUSY AdS vacuum have identical tensions! This is why, within the approximations made, the non-SUSY AdS vacua are marginally stable with respect to bubble nucleation.

Interestingly, a recent investigation \cite{Marchesano:2021ycx} of the non-SUSY AdS$_4$ vacua has revealed that the domain walls corresponding to wrapped D8 branes can induce the required instabilities, but this is only visible beyond the smeared approximation. What really happens is that the thin wall approximation used in \cite{Narayan:2010em} is not entirely valid since the moduli are light enough to invalidate it. Then any correction to the theory can make the marginal decay tip over to one side or another. Marchesano et al showed \cite{Marchesano:2021ycx} that this indeed happens exactly as predicted by the Swampland conjecture \cite{Ooguri:2016pdq}. The details require going beyond the smeared approximation and regarding the wrapped D8 properly as a D8-D6 boundstate as predicted by the Freed-Witten anomalies discussed earlier.

One would expect the same to be happening here since the wrapped D8 branes are equally present in our setup. To analyse this, the backreaction of the orientifolds in our model should be computed. 
We leave that for the future but expect that the difference between SUSY and fake SUSY is something that has consequences for stability only when going beyond the leading lower-dimensional SUGRA approximation.

\section{Holography} \label{sec:conlon}
A way to check the consistency of the scale-separated AdS$_4$ and AdS$_3$ vacua, and therefore the Strong AdS Distance Conjecture \cite{Lust:2019zwm}, is to rule in or out the would-be dual CFTs. These CFTs should have a parametric large central charge $c$ and a sparse spectrum of low-lying single-trace operators which is separated from heavier operators by a large gap. Moreover, if the AdS vacua are to be used for uplifting to de Sitter vacua, the dual CFTs should be `dead-end', meaning that there are no marginal or relevant deformations.
 There are no CFTs known with these properties. In fact, there is even only one dead-end CFT known: the 2D Monster CFT at $c=24$ \cite{Frenkel3256}. CFTs with a sparse spectrum as described above are also difficult to bootstrap, because the standard crossing symmetry requirements are trivially satisfied at large $c$: such CFTs have a large $N$ parameter and there is a one-to-one correspondence between EFTs in AdS and CFT data in a 1/N expansion \cite{Heemskerk_2009}. For 2D CFTs, the enhanced symmetry algebra and the modular bootstrap (f.e. \cite{Hellerman_2011}, \cite{Friedan_2013}, \cite{collier2016modular})  might play to one's advantage. However, with all current modular bootstrap bounds scaling with the central charge $c$, it is still a challenge to put this to use on the scale-separated spectra.
 
Finally, we notice that for CFT's above dimension two, the recent work \cite{Collins:2022nux} shows that in a large class of flux compactifications arising from branes probing singularities of internal manifolds there is a lower bound in the diameter of the internal space in AdS units, which itself translates to an upper bound in the gap of massive spin 2 operators, corresponding to graviton KK modes in the present context. If such a bound is true generally, scale-separated AdS vacua are ruled out.

\subsection{CFT duals for AdS$_3$ vacua}
Below, we present some basic properties of the would-be CFT$_2$ duals to the scale-separated AdS$_3$ vacua of \cite{Farakos:2020phe}.

\paragraph{Symmetry algebra.}

Let us start by giving the symmetry algebra of the dual 2-dimensional conformal field theory. Since the supergravity theory has 2 real supercharges, the CFT must have $N=(1,0)$ or $N=(0,1)$ supersymmetry, depending on whether O6- or $\overline{\text{O}}$6-planes were used in the compactification. If the G2-space has $b_2 \neq 0$, there are additionally vector fields arising from the reduction of the RR 3-form along the 2-cycles, which have a Chern-Simons mass. Following \cite{Montero:2016tif}, these are dual to (anti-)holomorphic currents $J^a(z)=\sum_n j_n^a z^{(-n+1)}$ for a positive Chern-Simons mass. \\

\paragraph{Central charge} 
The central charge is related to the AdS scale $L$ by
\begin{equation}
    c = \dfrac{3L}{2G_3},
\end{equation}
with gravitational constant $G_3 = (16\pi)^{-1}l_p$ \footnote{We use the Brown-Henneaux convention $S = \dfrac{1}{16 \pi G_N}\int (R-2 \Lambda)$, and $\Lambda=-1/L^2$.} and
where $l_p$ is the 3-dimensional Planck scale.\\
The AdS scale is given by
\begin{equation}
    L = \abs{V_0/2}^{-1/2} l_p,
\end{equation}
where the vev of the potential is \cite{Farakos:2020phe}
\begin{equation}
    \abs{V_0} = 0.070565 \cdot \dfrac{m^4h^6}{f^8},
\end{equation}
so that
\begin{equation}
    c =   7.9809 \cdot(16\pi) \dfrac{f^4}{m^2 h^3}.
\end{equation}

\paragraph{Operator spectrum.}

To compute the conformal dimensions of the operators dual to the 8 scalars originating from 7 metric fluctuations and the dilaton, we need to be careful with normalisations and conventions. We will follow \cite{Farakos:2020phe} and then the relevant part of the Lagrangian is given by:
\begin{equation}\label{delta}
e^{-1}\mathcal{L} = \frac{1}{2}\mathcal{R} -\frac{1}{4}\delta_{ab}\partial\phi^a\partial\phi^b  + \frac{1}{L^2} - \frac{1}{4}\sum_a m_a^2\phi_a^2\,,
\end{equation}
where $a,b$ runs from $1, \ldots, 8$. We only displayed the potential to quadratic order and in a basis that diagonalises the mass matrix and kinetic terms. In these conventions the operator dimensions in the CFT are:
\begin{equation}
    \Delta_a = 1+ \sqrt{1+m_a^2L^2}.
\end{equation}
The kinetic terms in the Lagrangian are almost of the canonical form \eqref{delta}, but not quite:
\begin{equation}
    e^{-1} \mathcal{L}_{kin} = -\dfrac{1}{4}(\partial x)^2-\dfrac{1}{4} (\partial y)^2 -\dfrac{1}{4} \dfrac{(\partial \tilde{s}^i)^2}{(\tilde{s}^i)^2}\,,
\end{equation}
To go to canonically normalised moduli \eqref{delta}, we define $S^i \equiv \log \tilde{s}^i$. The cosmological constant $-1/L^2$ equals the value $-4P^2$ for the extremum of the real (or fake) $P$ for the (fake) SUSY vacuum. The masses in the Lagrangian above are obtained from diagonalising the Hessian of $8(\delta^{ab}P_aP_b-P^2)$, where again the indices are related to the canonically normalised scalars defined above. We then find:
\begin{equation}
   m_a^2L^2 = 2\cdot(53.99, 5.54, 3.53, 1.69, 1.69, 1.69, 1.69, 1.69), 
\end{equation}
such that:
\begin{equation}
    \Delta_a = (11.44, 4.48, 3.84, 3.09,3.09,3.09,3.09,3.09 ).
\end{equation}
These are the light single trace operators in the CFT and our main observation is that they are not integer, unlike the operators in the CFT dual to the DGKT vacua \cite{Conlon:2021cjk}.

The Kaluza-Klein modes will correspond to heavy operators. The vev of the volume in string frame is given by
\begin{equation}
    V_S = 1.300079 \cdot m^{-3/4} h^{-1} f^{7/4},
\end{equation}
and with this the masses of the Kaluza-Klein modes are 
\begin{equation}
    m_{KK} =  V_S^{-1/7} l_s^{-1} =  V_S^{-1/7}  (l_{p} \cdot V_S/e^{2\phi})^{-1} =5.160344 m^{5/14}h^{22/7}f^{-7/2}l_p^{-1}
\end{equation}
where $l_s$ is the string scale,
and thus
\begin{equation}
    \Delta_{KK} \sim f^{1/2}.
\end{equation}
Other states in the CFT can come from wrapping D-branes on $p$-cycles in AdS. The only options that are not projected out by the orientifold are D2-branes wrapped on 2-cycles, if the twisted sector contains such cycles. Their mass is given by
\begin{equation}
    m_{D_2} \sim e^{-\phi} l_s^{-1} V_S^{2/7} \sim f^{-2} h^{12/7} m^{2/7}l_p^{-1},
\end{equation}
and so the dimension of the dual operator scales with the F4-flux like
\begin{equation}
    \Delta_{D_2} \sim f^2.
\end{equation}
A summary of the spectrum is shown in Table \ref{table:spectrum}.
\begin{table}[h!]
    \begin{center}
    \begin{tabular}{ | l | l | }
    \hline
    \textbf{Central charge} & $c \sim f^4$\\
    \hline 
    \textbf{Operator dimensions} &  \\
    \hline
    \emph{light} & \\
    \hline
    graviton, gravitino, moduli & $\Delta\sim \mathcal{O}(1)$\\
    \hline
    \emph{medium} & \\
    \hline
    KK modes & $\Delta_{KK} \sim f^{1/2}$\\
    \hline
    wrapped D2 branes & $\Delta_{D2} \sim f^2$\\
    \hline
    \emph{heavy} & \\
    \hline
    BTZ black hole & $\Delta_{BH} \gtrsim  f^4$\\
    \hline
    \end{tabular}
\end{center}
\caption{The spectrum.}\label{table:spectrum}
\end{table}

\paragraph{Sign of anomalous dimensions?}

Conlon and Revello \cite{Conlon:2020wmc} suggested that certain anomalous dimensions of double trace operators should be negative for having a consistent CFT dual, along the lines of ideas presented in \cite{Nachtmann:1973mr}, \cite{Hartman_2015}. Especially for double trace operators made from an operator dual to an axion \emph{and} a saxion this seemed almost identical to having an axion decay constant that does not become trans Planckian. But for operators that are different one can actually violate this \cite{Conlon:2021cjk}.  In our 3D model we have no axions and hence this suggestion cannot be tested.

\subsection{Scaling dimensions for CFT duals of AdS$_4$ vacua}\label{Timmsection}
Given that we do not find integer dimensions for the dual operators of the AdS$_3$ vacua in the previous subsection, we revisit here the status of AdS$_4$ vacua. In \cite{Conlon:2021cjk} the authors showed that for the specific compactification of massive type IIA on $T^6/\mathbb{Z}_3 \times \mathbb{Z}_3$, with four complex light fields, all dual operators have integer dimensions. In particular, there are three 2-cycle volume moduli $v_i$, with overall volume $vol_6\propto v_1 v_2 v_3$, and the dilaton, as well as three $B_2$-axions and one $C_3$-axion. Given the highly symmetric setting with a rather simple relation between the 2-cycle volumina $v_i$ and the overall volume $vol_6$ and an actual exchange symmetry between three of the four complex moduli, one might wonder whether this example is special. Hence this motivates us to study a broader class of massive type IIA flux compactifications and calculate the operator dimensions for the putative CFT$_3$ duals. Another simplification arose for $T^6/\mathbb{Z}_3 \times \mathbb{Z}_3$ since it has no complex structure moduli because $h^{2,1}=0$. The models we will study will also include complex structure moduli and we will explicitly verify their expected dual operator dimensions discussed in \cite{Conlon:2021cjk}, based on results previously obtained in \cite{Conlon:2006tq}.

The class of models we will discuss first are abelian toroidal orbifolds and in particular we will study the examples listed in table 5 of \cite{Flauger:2008ad}. We will, following \cite{Flauger:2008ad}, restrict to bulk moduli only and hence there are 11 different classes of models (some of which are subclasses of others and one class contains $T^6/\mathbb{Z}_3 \times \mathbb{Z}_3$). We studied massive type IIA in the presence of O6-planes, H-flux and RR-fluxes and found the supersymmetric AdS$_4$ vacua for each of these models. We then calculated the masses for the light fields, which are \\

\begin{enumerate}
\item $h^{1,1}_-$ 2-cycle volumes and their axionic partners from $B_2$,
\item the dilaton and its axionic partner from $C_3$,
\item $h^{2,1}$ complex structure moduli and their axionic partners from $C_3$.
\end{enumerate}

In practice the K\"ahler moduli arise from integrating $J_c=B_2 + \rmi J$ over the $h^{1,1}_-$ 2-cycles that are odd under the O6-plane involution. The complex structure moduli arise from integrating $\Omega_c = C_3+\rmi e^{-\phi} Re(\Omega)$ over the 1+$h^{2,1}$ even 3-cycles. The superpotential involves via $W \supset \int H\w \Omega_c$ only one linear combination of the complexified complex structure moduli, hence the split above into 2. and 3.. The $C_3$ axions do not appear in the K\"ahler potential and this means that only one linear combination of the $C_3$-axions appears in the scalar potential at all, namely $\int H\w C_3$. All the other linear combinations of $C_3$ axions are flat directions with $m^2=0$. This then fixes the masses of the corresponding saxionic complex structure moduli to $m^2 =-2/3 V_{min} = -2/R_{AdS}^2$, where $V_{min}$ is the value of the scalar potential in the minimum and $R_{AdS}$ is the corresponding radius of the AdS space \cite{Conlon:2006tq, Conlon:2021cjk}. The corresponding dual operator dimension $\Delta(\Delta-3)=m^2 R_{AdS}^2$ is then $\Delta=1$ or $\Delta=2$.

We find for all the supersymmetric AdS$_4$ vacua in our eleven classes of models that there is one universal complex direction that involves the complexified dilaton and the overall volume. This complex directions has masses squared such that the dual integer operator dimensions $\Delta=1/2 \left( 3 + \sqrt{9 + 4 m^2 R_{AdS}^2}\right)$ are 10 and 11 for the saxion and axion, respectively. We will refer to it as the dilaton direction 2. but we stress that it is actually also involving the K\"ahler moduli in 1. above via the overall volume. We can then summarize the six different dual operator dimensions corresponding to light fields in all the 11 classes of models as shown in table \ref{tab:SUSY}

\begin{table}[h]
\begin{center}
\begin{tabular}{|c|c|}\hline
Modulus & Operator dimension $\Delta$\\\hline\hline
1. $h^{1,1}_-$ saxionic K\"ahler moduli from $J$ &  6\\
1. $h^{1,1}_-$ axionic K\"ahler moduli from $B_2$ &  5\\\hline
2. The dilaton direction & 10\\
2. The $C_3$-axion appearing in $W$ & 11\\ \hline
3. $h^{2,1}$ saxionic complex structure moduli from $Re(\Omega)$ &  1 or 2\\
3. $h^{2,1}$ massless $C_3$-axions &  3\\\hline
\end{tabular}
\caption{Summary of integer operator dimension of a putative CFT$_3$ dual for generic supersymmetric DGKT type AdS$_4$ vacua.}\label{tab:SUSY}
\end{center}
\end{table}

Note that there is an expected degeneracy among the $h^{2,1}$ complex structure moduli that all have massless axionic partners \cite{Conlon:2006tq}. However, it is highly surprising that there is a similar degeneracy among all the $h^{1,1}_-$ K\"ahler moduli. The original model in DGKT \cite{DeWolfe:2005uu} had an exchange symmetry among these moduli and therefore it was more natural to find the same masses and dual conformal operator dimensions. However, we have now studied also models, like $T^6/\mathbb{Z}_{6-I}$, with for example four K\"ahler moduli and a volume $vol_6 \propto v_1 v_2 v_3 - v_1 v_4 v_4$. \footnote{We also studied the way more complicated model $T^6/\mathbb{Z}_3$ model with six K\"ahler moduli and $vol_6 \propto v_1 (v_2 v_3 - v_4^2) + v_2 v_5^2+ v_3 v_6^2 + 2 v_4 v_5 v_6$.} These models have no such exchange symmetry but the resulting masses are nevertheless all degenerate. This hints at a general deeper reason and it would be interesting to try to prove this in full generality. 

The supersymmetric DGKT AdS vacua have closely related non-supersymmetric AdS vacua that are related to the supersymmetric vacua by sign flips of $F_4$ flux quanta \cite{DeWolfe:2005uu}. The original $T^6/\mathbb{Z}_3 \times \mathbb{Z}_3$ model had three $F_4$ flux quanta and corresponding non-supersymmetric vacua that were obtained by flipping the sign of any one, two or all three of these quanta. Whenever one would flip the sign of one or three $F_4$- flux quanta, the masses of the axionic directions would change so that the conformal scaling dimensions of their dual operators are 8 for the $B_2$ axions and 1 or 2 for the $C_3$-axion. We find that, while more complicated models do not have non-supersymmetric vacua for a sign flip of any choice of $F_4$-flux quanta, they have non-supersymmetric vacua that can be obtained by flipping the sign of all $F_4$-flux quanta. In that case we find in all the abelian toroidal orbifold models the following dual operator dimensions shown in table \ref{tab:nonSUSY}.

\begin{table}[h]
\begin{center}
\begin{tabular}{|c|c|}\hline
Modulus & Operator dimension $\Delta$\\\hline\hline
1. $h^{1,1}_-$ saxionic K\"ahler moduli from $J$ &  6\\
1. $h^{1,1}_-$ axionic K\"ahler moduli from $B_2$ &  8\\\hline
2. The dilaton direction & 10\\
2. The $C_3$-axion appearing in $W$ & 1 or 2\\ \hline
3. $h^{2,1}$ saxionic complex structure moduli from $Re(\Omega)$ &  1 or 2\\
3. $h^{2,1}$ massless $C_3$-axions &  3\\\hline
\end{tabular}
\caption{Summary of integer operator dimension of a putative CFT$_3$ dual for non-supersymmetric DGKT type AdS$_4$ vacua obtained by flipping the signs of $F_4$-flux quanta. Note that all non-flat axionic directions have different masses now.}\label{tab:nonSUSY}
\end{center}
\end{table}

We studied a few more non-supersymmetric AdS vacua and found that the dual operator dimension were always given by either the dimensions listed in the supersymmetric table \ref{tab:SUSY} or the non-supersymmetric table \ref{tab:nonSUSY}. For example, for the $T^6/\mathbb{Z}_{6-I}$ case mentioned above, there are four $F_4$ flux quanta. Non-supersymmetric AdS vacua exist for a sign flip of the first one (or all of them) with the dimensions given in table \ref{tab:nonSUSY}. However, there are also non-supersymmetric AdS vacua obtained by a sign flip of the second, third and fourth flux quanta with the dimensions given in table \ref{tab:SUSY}. Again it would be interesting to understand this better and in full generality.

Given that the DGKT setup with Ricci flat Calabi-Yau manifolds is a special subclass within the compactifications on more general $SU(3)\times SU(3)$ manifolds, it is naturally to ask whether the above generalizes. There is a somewhat trivial class of non-Ricci-flat compactifications that is related to the DGKT setup by two T-dualities \cite{Banks:2006hg, Cribiori:2021djm}. Given the relation via T-duality to the original DGKT setup or the models we studied above in this subsection, all these cases trivially give rise to integer conformal dimension for the putative CFT$_3$ duals. Note however, that these models can be potentially strongly coupled and therefore could arise from M-theory compactifications \cite{Banks:2006hg, Cribiori:2021djm}, which might or might not make the search for a CFT dual more tractable.

Next we looked at models that contain metric fluxes, i.e. have non-trivial curvature, and are not T-dual to models that are Ricci flat. For example, one interesting models is discussed in \cite{Font:2019uva} (see also \cite{Camara:2005dc}). In the model discussed in subsection 2.1 in \cite{Font:2019uva} the authors find, for a model with metric fluxes but without mass parameters, masses squared that are (see their equation (2.14))
\begin{equation}
m^2 R_{AdS}^2= \{ 18, 22/9, -2, 10, -8/9, 0\}\,.
\end{equation}
While these are not integers as was the case before, it seems equally surprising that they are rational numbers and so are the dual operator dimensions
\be
\Delta = \{ 6, 11/3,1 \text{ or } 2, 5, 8/3, 3\}\,.
\ee
So, it seems there might be a larger class of models that has potentially rational dual operator dimensions and the Calabi-Yau compacitifications are a subclass with integer dimensions.\footnote{The paper \cite{Font:2019uva} discusses one further model in subsection 2.2 that has irrational dimensions for the dual operators (cf. their equation (2.20).} Interestingly this vacuum is not scale separated \cite{Font:2019uva} and so one can wonder whether in 4D the integer nature of the dual conformal dimensions is related to scale separation.

Lastly, there are in the literature geometric type IIB compactifications that allow for full moduli stabilization at tree-level with fluxes and O5/O7-planes \cite{Caviezel:2009tu, Petrini:2013ika}. These examples are related by a single T-duality to type IIA flux compactifications with geometric and non-geometric fluxes and have unconstrained flux parameters. However, all existing models have no limit in which all cycle volumes become large and the string coupling weak, as was recently clarified in \cite{Cribiori:2021djm}. In this class of compactifications we studied the supersymmetric AdS$_4$ vacuum that is given in subsection 4.1 of \cite{Caviezel:2009tu}. We found that the mass matrix is an extremely complicated expression with square-roots and the dual operator dimensions are not integers (and are most likely also not rational numbers).

\section{Conclusion}
In this note we have verified that the scale separated AdS$_3$ vacua of \cite{Farakos:2020phe} do not pass the refined AdS Distance Conjecture of \cite{Buratti:2020kda} because the required discrete higher form symmetries are absent, despite the very close analogy with the DGKT vacua \cite{DeWolfe:2005uu} that do have them. We furthermore verified that the single trace operators dual to the light scalars do not have integer dimensions, also in contrast with the DGKT vacua \cite{Conlon:2021cjk}. Given these observation we revisited different types of AdS$_4$ vacua of DGKT type, as well as generalization thereof. The goal was to get an understanding of whether the particular example of $T^6/\mathbb{Z}_3 \times \mathbb{Z}_3$, studied in \cite{DeWolfe:2005uu, Conlon:2021cjk}, is special or whether integer dimensions for dual operators arise more broadly. We find indeed that for all orbifold models we checked that feature scale separation, the conformal dimensions are always integers. Therefore we studied a few more general compactifications involving metric fluxes, i.e. non-flat geometries, and find that this feature does generically not persist but then also scale separation was absent. This suggest that there might be a link between integer operator dimensions, discrete higher form symmetries and scale separation that deserves further study. 

We furthermore verified that the stability arguments of \cite{Narayan:2010em} for non-SUSY DGKT vacua can be readily extended to the 3D case, indicating that one does not find an instability at tree level for the non-SUSY vacua.

Our observations related to AdS$_3$ vacua might need to be refined if one goes beyond the approximations made. In particular, one would expect the D8 branes to induce non-perturbative instabilities for the non-SUSY vacua once backreaction is taken into account as in \cite{Marchesano:2021ycx}. The absence of discrete symmetries and integer dimensions for AdS$_3$ could potentially be affected if we study regular instead of singular G2 manifolds. This could be achieved by extending our orbifold symmetries with the appropriate shifts such that they are part of the Joyce class that allows a desingularisation, which introduces an extra twisted sector. One could wonder whether mass mixing with the twisted sector could make the operator dimensions integer, but we believe this is unlikely, simply from analogy with the twisted sector in DGKT. The scalars in the untwisted sector really provide the universal sector that is most sensitive to the large flux limit and all effects of a twisted subsector seem subleading. The same applies to the analysis of the discrete symmetries. It would nonetheless be interesting to construct AdS$_3$ vacua on regular G2 spaces and verify this explicitly. 

It also seems worthwhile to study AdS$_4$ vacua in more detail. Can we proof in full generality that all CY$_3$ flux compactifications of massive type IIA with smeared O6-planes have light moduli masses that lead to dual operators with only integer dimensions? Is it possible to understand when exactly and why AdS$_4$ vacua give rise to such dual CFT$_3$ with integer operator dimensions? We leave these enticing questions for the future.

\section*{Acknowledgments}
We like to thank Nikolay Bobev, Joe Conlon, Fotis Farakos, George Tringas and Vincent Van Hemelryck for useful discussions. This work is supported by the KU Leuven C1 grant ZKD1118C16/16/005. FA acknowledges the Clarendon Fund Scholarship in partnership with the Scatcherd European Scholarship, Saven European Scholarship and the Hertford College Peter Howard Scholarship. The work of TW is supported in part by the NSF grant PHY-2013988.

\small{\bibliography{refs}}
\bibliographystyle{utphys}
\end{document}